\documentclass[sigconf, nonacm]{acmart}

\AtBeginDocument{%
  }

\usepackage{amsmath,amsfonts}
\usepackage{algorithmic}
\usepackage{graphicx}
\usepackage{subcaption}
\usepackage{textcomp}
\usepackage{xcolor}
\usepackage{xspace}
\usepackage{fontawesome}
\usepackage{multirow}
\usepackage{rotating}
\usepackage{tikz}
\usepackage{soul}
\usepackage{makecell}
\usepackage{url}
\usepackage{hyperref}
\usepackage{enumitem}
\usepackage{booktabs}
\usepackage{fontawesome}
\usepackage{pifont}
\usepackage[most]{tcolorbox}
\usepackage{colortbl}
\usepackage{adjustbox}
\usepackage{listings}
\usepackage[utf8]{inputenc}
\usepackage{textcomp}

\newcommand{\lstbg}[3][0pt]{{\fboxsep#1\colorbox{#2}{\strut #3}}}
\lstdefinelanguage{diff}{
  basicstyle=\ttfamily\scriptsize,
  morecomment=[f][\lstbg{red!20}]-,
  morecomment=[f][\lstbg{green!20}]+,
  morecomment=[f][\textit]{@@},
}

\newtcolorbox{promptbox}{colback=white, arc=0.5mm, top=1mm, bottom=1mm, left=1mm, right=1mm, title=System prompt used for generation}

\newtcolorbox{resultbox}{
  colback=white,
  arc=0.5mm,
  top=1mm,
  bottom=1mm,
  left=1mm,
  right=1mm,
  fontupper=\normalfont
}

\newcommand{\ie}{\emph{i.e.,}\xspace}
\newcommand{\gpt}{\emph{GPT-4o}\xspace}
\newcommand{\eg}{\emph{e.g.,}\xspace}

\newcommand{\etal}{\emph{et~al.}\xspace}

\newcommand{\tabref}[1]{Table~\ref{#1}\xspace}

\newcommand{\classeval}{ClassEval\xspace}

\definecolor{lightlightgrey}{RGB}{233, 236, 239}

\newboolean{showcomments}
\setboolean{showcomments}{true}

\ifthenelse{\boolean{showcomments}}
  {\newcommand{\nb}[2]{
    \fbox{\bfseries\sffamily\scriptsize#1}
    {\sf\small$\blacktriangleright$\textit{#2}$\blacktriangleleft$}
   }
  }
  {\newcommand{\nb}[2]{}
  }

\setcopyright{none} 
\renewcommand\footnotetextcopyrightpermission[1]{}

{\begin{list}{}%
   {\leftmargin=#1 \rightmargin=#1}%
   \item\relax}
{\end{list}}

\begin{document}

\title[From Human to Machine Refactoring: Assessing GPT-4’s Impact on Python Class Quality and Readability]{From Human to Machine Refactoring: Assessing GPT-4’s Impact on Python Class Quality and Readability}

\author{Alessandro Midolo}
\email{alessandro.midolo@unict.it}
\affiliation{%
  \institution{Dipartimento di Matematica e Informatica, University of Catania}
  \city{Catania}
  \country{Italy}
}

\author{Emiliano Tramontana}
\email{tramontana@dmi.unict.it}
\affiliation{%
  \institution{Dipartimento di Matematica e Informatica, University of Catania}
  \city{Catania}
  \country{Italy}
}

\author{Massimiliano Di Penta}
\email{dipenta@unisannio.it}
\affiliation{%
  \institution{University of Sannio}
  \city{Benevento}
  \country{Italy}
}

\begin{abstract}
Refactoring is a software engineering practice that aims to improve code quality without altering program behavior. Although automated refactoring tools have been extensively studied, their practical applicability remains limited. Recent advances in Large Language Models (LLMs) have introduced new opportunities for automated code refactoring.
The evaluation of such an LLM-driven approach, however, leaves unanswered questions about its effects on code quality.
In this paper, we present a comprehensive empirical study on LLM-driven refactoring using \gpt, applied to 100 Python classes from the ClassEval benchmark. Unlike prior work, our study explores a wide range of class-level refactorings inspired by Fowler’s catalog and evaluates their effects from three complementary perspectives: (i) behavioral correctness, verified through unit tests; (ii) code quality, assessed via Pylint, Flake8, and SonarCloud; and (iii) readability, measured using a state-of-the-art readability tool. Our findings show that \gpt generally produces behavior-preserving refactorings that reduce code smells and improve quality metrics, albeit at the cost of decreased readability. 
Our results provide new evidence on the capabilities and limitations of LLMs in automated software refactoring, highlighting directions for integrating LLMs into practical refactoring workflows.
\end{abstract}
\maketitle

\section{Introduction}
\label{sec:introduction}

Refactoring is defined as “the process of changing a software system in such a way that it does not alter the external behavior of the code yet improves its internal structure”~\cite{fowler2002refactoring}. The goal of refactoring is to enhance software quality---e.g., understandability and maintainability---by restructuring source code without affecting its behavior. Over the years, it has become an essential practice in software evolution and maintenance, allowing developers to manage technical debt, increase readability, and extend the lifespan of complex software systems. However, performing refactoring activities manually can be tedious and error-prone, particularly in large codebases where unintended behavioral changes may arise. This has motivated the development of numerous automated refactoring tools and techniques~\cite{midolo2021refactoring, kurbatova2020recommendation, niu2023rat, tsantalis2022refactoring, ellis2023operation}, which aim to assist developers in executing safe and systematic code transformations.

In recent years, the rise of Large Language Models (LLMs) has opened the possibility of better software engineering automation. This includes automated software refactoring. LLMs trained on vast corpora of source code exhibit strong code understanding and generation capabilities, making them promising candidates for automating or assisting in complex software transformations. Several studies have explored this potential by combining LLMs with static or dynamic program analysis to enhance automated refactoring activities. E.g., EM-Assist~\cite{pomian2024emassist} and iSMELL~\cite{wu2024ismell} integrate LLM suggestions with static analysis to apply behavior preserving Extract Method technique and for code-smell refactoring, respectively. Other approaches employ LLMs to generate synthetic test programs to expose refactoring opportunities~\cite{cordeiro2024empirical}, detect code smells~\cite{gheyi2025evaluating}, or recommend specific transformations such as the Move Method technique~\cite{zhang2025move} and test-smell refactoring~\cite{gao2025automated}. Empirical investigations by Shirafuji \etal~\cite{shirafuji2023refactoring} and Cordeiro \etal~\cite{cordeiro2024empirical} have evaluated the general refactoring capabilities of modern LLMs, while Midolo \etal~\cite{midolo2025automated} and Liu \etal~\cite{liu2024empiricalstudypotentialllms, liu2025exploring} examined their effectiveness in realistic or style-oriented settings.

While these studies show the promise of LLM-assisted refactoring, they exhibit notable limitations. Most efforts focus on correctness, removing code smells, or executing single refactoring types. Only a few works have quantitatively assessed whether LLM-generated refactorings truly improve code quality, as measured by static analysis metrics or readability measures. Furthermore, issues such as hallucination, insufficient contextual reasoning, and the lack of systematic validation remain open challenges~\cite{cordeiro2025llm}.

To address these limitations, we propose an empirical study on LLM-driven code refactoring using \gpt. Our study builds upon the \classeval~\cite{xueying2023classeval} benchmark, which provides 100 class-level Python tasks with canonical solutions and associated test suites. We use the canonical solutions as input to \gpt, prompting it to apply a wide set of class-level refactorings derived from Fowler’s catalog~\cite{fowler2002refactoring}. Unlike prior works that focus on specific refactoring types, our study encompasses a diverse set of transformations. We then evaluate the refactored code under three complementary perspectives: (i) correctness, by executing the \classeval test suites; (ii) code quality, through static analysis with Pylint~\cite{pylint}, Flake8~\cite{flake8}, and SonarCloud~\cite{sonarcloud}; and (iii) readability, using the metric proposed by Scalabrino \etal~\cite{scalabrino2018comprehensive}.

Our findings reveal that \gpt tends to produce behavior-preserving refactorings that generally enhance the overall quality of the code. In particular, the static analysis results show a reduction in the number of code smells and warnings after the refactoring. However, while code quality
(i.e.\ cyclomatic complexity) 
improves, readability tends to degrade, suggesting that \gpt prioritizes structural improvements over code clarity.

Our main contributions are as follows.
\begin{itemize}
    \item A comprehensive empirical evaluation of LLM-based refactoring at the class level, encompassing a broad set of refactoring types derived from Fowler’s catalog.
    \item A multi-perspective assessment framework combining correctness, code quality, and readability analysis through standardized benchmarks and static analysis tools.
    \item An open replication package including datasets, prompts, and analysis scripts to support transparency and reproducibility.
\end{itemize}

The remainder of this paper is structured as follows.
Section~\ref{sec:related} reviews related work.
Section~\ref{sec:design} presents the design of our empirical study.
Section~\ref{sec:results} reports and discusses our findings.
Section~\ref{sec:threats} discusses threats to validity, and Section~\ref{sec:conclusion} concludes the paper and outlines future research directions.

\section{Related Work} \label{sec:related}

Recent advances in large language models (LLMs) have spurred extensive research on their application to software maintenance, particularly automated code refactoring and quality improvement. Prior studies investigate how LLMs can generate, recommend, or validate refactorings, often combining them with static analysis, traditional refactoring tools, or employing hybrid prompting techniques. In this section, we first review related work on automated code refactoring supported by LLMs, then discuss broader efforts applying LLMs to software engineering tasks that inform our approach and evaluation design.

\subsection{Automated Code Refactoring}

Recent studies highlighted the growing role of LLM in automated software refactoring. Several works integrated LLMs with traditional program analysis to improve the precision and safety of automated transformations. EM-Assist~\cite{pomian2024next,pomian2024emassist} augments the classic Extract Method refactoring technique by combining iterative LLM suggestions with IDE static-analysis filters and safe engine execution.
Similarly,
iSMELL~\cite{wu2024ismell} mixes multiple static analyzers and LLMs in a Mixture-of-Experts architecture for code-smell detection and refactoring. Dong \etal~\cite{dong2025chatgpt} leveraged ChatGPT to generate synthetic test programs to expose bugs in refactoring engines, while Gheyí \etal~\cite{gheyi2025evaluating} evaluated small language models for detecting refactoring bugs. These works confirm the value of static analysis and LLM synergy. However, none systematically assesses the quality of LLM-generated refactorings through static analysis metrics as we do.

Other approaches focused on recommending or performing specific refactorings. MoveRec~\cite{zhang2025move} combines structural metrics and LLM-derived semantic features to predict move-method refactorings. UTRefactor~\cite{gao2025automated} targets unit test smells using a chain-of-thought LLM pipeline, and Baumgartner \etal~\cite{baumgartner2024ai} tackle data-clump smells with a hybrid AST/LLM pipeline. Midolo \etal\cite{midolo2025automated} conducted a replication study to assess whether GPT is able to refactor Python code toward a Pythonic style, comparing the results with a static tool. Shirafuji \etal~\cite{shirafuji2023refactoring} and Cordeiro \etal~\cite{cordeiro2024empirical} empirically studied the capacity of LLMs (e.g., GPT-3.5, StarCoder2) to perform general code refactoring, while Liu \etal~\cite{liu2024empiricalstudypotentialllms,liu2025exploring} and Robredo \etal~\cite{robredo2025what} investigated LLM effectiveness and motivations in real-world refactorings. Rajendran \etal~\cite{rajendran2025multi} proposed a multi-agent LLM framework to balance competing refactoring goals. These contributions demonstrate the promise of LLM-assisted refactoring, yet they primarily evaluate correctness or smell reduction, rather than quantitatively measuring code quality improvement.

Prompt engineering and hybrid strategies have also been explored to enhance LLM-based refactoring. White \etal~\cite{white2024chatgpt} categorize reusable prompt patterns, Alomar \etal~\cite{alomar2024refactor} analyze how developers and ChatGPT negotiate refactoring intents, and Ishizue \etal~\cite{ishizue2024improved} integrate refactoring tools with GPT for improved program repair. Cordeiro \etal~\cite{cordeiro2025llm} survey challenges such as hallucination, lack of context, and missing safety guarantees, calling for stronger integration of LLMs with static analysis and validation—gaps our work directly addresses.

In contrast to these efforts, our approach goes beyond proposing or applying LLM-generated refactorings. We assess the quality of the performed refactorings using established static analysis tools and incorporate a readability score to evaluate whether the generated changes actually improve code readability—an aspect that has been largely overlooked in previous studies. Furthermore, we validate our findings using a well-established dataset, ensuring that our conclusions are based on a solid empirical foundation. This combination of systematic quality assessment, readability evaluation, and benchmark-based validation represents the key novelty of our work compared to the existing literature.

Previous work \cite{BavotaLPOP15} also analyzed the relationship between refactoring (in this case, performed by humans) and code quality. Results indicate that there is rarely a clear relationship between refactoring and metrics' improvement, or smell removal.

\subsection{LLMs for Software Engineering}

Several studies have explored the use of LLMs in software engineering. A comprehensive overview can be found in existing systematic literature reviews~\cite{xinyi2024large, fan2023large}. Here, we highlight representative works most relevant to our context.

Depalma \etal~\cite{depalma2024exploring} evaluate the ability of LLMs to refactor Java code, with a focus on loop optimizations and overall code quality improvements. Chavan \etal~\cite{chavan2024analyzing} examine developer interactions with ChatGPT on GitHub and Hacker News, identifying how it supports code refactoring activities. Guo \etal~\cite{guo2024exploring} assess ChatGPT’s effectiveness in automated code review, showing it outperforms the CodeReviewer tool on standard benchmarks. Siddiq \etal~\cite{siddiq2024quality} analyze the DevGPT dataset to study the quality of ChatGPT-generated Python and Java code, highlighting recurring issues such as security vulnerabilities and insufficient documentation. Tufano \etal~\cite{tufano2024unveiling} mine GitHub references to ChatGPT to build a taxonomy of 45 developer tasks automated with LLMs, notably including refactoring. Poldrack \etal~\cite{poldrack2023ai} investigate GPT-4’s ability to refactor code and improve quality metrics such as maintainability, cyclomatic complexity, and compliance with coding standards. Finally, Noever \etal~\cite{noever2023chatbots} show how AI-driven code assistants can analyze and enhance historically significant codebases, clarifying obfuscated code and improving performance.

These works collectively demonstrate the growing impact of LLMs on code quality and refactoring, providing the foundation for our study.

\section{Study Design} \label{sec:design}
The \emph{goal} of this study is to evaluate the performance of LLMs,
specifically \gpt, in refactoring Python source code. 
We focus on the model’s ability to automatically refactor Python classes according to the techniques described by Fowler~\cite{fowler2002refactoring}. 
The \emph{quality focus} is on the one hand the correctness of the refactored code, and, on the other hand, its quality and readability, as measured by code metrics.

Our work addresses three research questions.

\begin{itemize}
    \item \textbf{RQ\textsubscript{1}} Can \gpt accurately refactor Python Classes while preserving the behavior of the generated code? This question investigates whether the refactored code remains functionally equivalent to the original. We assess this by executing the original test suite on the refactored code; any failing test indicates a behavioral change.
    \item \textbf{RQ\textsubscript{2}} 
    How does the refactored code by \gpt affect code quality metrics, design issues, and static code analysis tool violations? This question examines whether \gpt’s refactorings improve the internal structure of the code. We compare the results of static analysis tools applied to the original and refactored code to identify changes in code quality.
    \item \textbf{RQ\textsubscript{3}} To what extent the refactoring proposed by \gpt improve the readability of the code? This question explores whether \gpt’s transformations make the code easier to read and understand—a key, yet often overlooked, goal of refactoring. We assess readability differences between the original and refactored code.
\end{itemize}

By addressing these questions, this study provides a detailed assessment of GPT-4’s strengths and limitations in automated code refactoring, contributing to the broader discussion on the applicability of LLMs in software engineering. All data analyzed—including \gpt’s refactoring outputs, the associated metrics, and the results—are available in the paper’s public replication package~\cite{replication}.

\subsection{Dataset and Class-Level Refactorings} \label{sub:dataset}

For our study, we adopt ClassEval~\cite{xueying2023classeval}, a manually curated benchmark designed to evaluate LLMs on class-level code generation. In ClassEval, an entire Python class—often containing multiple interdependent methods and fields—must be generated from a provided skeleton and accompanying test suite.
Unlike method-level benchmarks such as HumanEval~\cite{chen2021evaluating}, ClassEval exposes LLMs to more realistic software-engineering challenges: handling cross-method dependencies, ensuring consistent field usage, and maintaining a coherent class structure. The dataset comprises 100 tasks (covering 410 methods), each providing both reference implementations and comprehensive test suites to verify functional correctness. This makes it appropriate for assessing refactoring tasks where preserving behavioral equivalence is critical.

\begin{table*}[htbp]
\centering
\small
\begin{tabular}{p{4cm}p{9cm}}
\hline
\textbf{Refactoring} & \textbf{Brief Description} \\
\hline
Extract Method & Move a fragment of code into a new method and call it from the original location. \\
Inline Method & Replace a method call with the method’s body when the method is no longer needed. \\
Replace Temp with Query & Substitute a temporary variable with a method that returns the same value. \\
Split Temporary Variable & Use separate variables when a single temporary is assigned multiple, unrelated values. \\
Remove Assignments to Parameters & Avoid changing parameter values inside a method; use local variables instead. \\
Replace Method with Method Object & Turn a complex method into its own class to handle local variables and simplify logic. \\
Encapsulate Field & Make a field private and provide accessor (getter/setter) methods. \\
Encapsulate Collection & Provide methods to access/modify a collection instead of exposing it directly. \\
Replace Magic Number with Symbolic Constant & Substitute literal numbers with named constants for clarity and maintainability. \\
Replace Data Value with Object & Replace a simple data item with an object to add behavior and meaning. \\
Self-Encapsulate Field & Access a field only through its own getter and setter, even within the same class. \\
Decompose Conditional & Extract complex conditional logic into separate methods for readability. \\
Consolidate Conditional Expression & Merge multiple conditionals that produce the same result into a single expression. \\
Consolidate Duplicate Conditional Fragments & Move identical code fragments outside of conditional branches. \\
Remove Control Flag & Eliminate a control flag variable by using break, return, or other clearer control structures. \\
Replace Nested Conditional with Guard Clauses & Use guard clauses to handle special cases early and reduce nested conditionals. \\
Introduce Null Object & Replace null references with an object that exhibits neutral behavior. \\
Add Parameter & Add a new parameter to a method to supply needed data. \\
Remove Parameter & Remove an unused or unnecessary parameter from a method. \\
Separate Query from Modifier & Split a method that both returns a value and changes state into two distinct methods. \\
Parameterize Method & Replace similar methods with a single one that uses parameters to vary behavior. \\
Preserve Whole Object & Pass an entire object instead of extracting and passing its individual fields. \\
Replace Parameter with Method Call & Remove a parameter whose value can be obtained by calling a method. \\
Replace Parameter with Explicit Methods & Replace a parameter used to select behavior with multiple methods, one for each behavior. \\
Rename Variable & Give a variable a more meaningful, descriptive name. \\
Introduce Explaining Variable & Assign the result of a complex expression to a well-named variable for clarity. \\
Inline Temp & Replace a temporary variable with the expression that computed its value. \\
Remove Dead Code & Delete code that is never executed or no longer needed. \\
Replace Loop with Pipeline & Replace loops with stream or pipeline operations (e.g., in functional programming). \\
\hline
\end{tabular}
\caption{Summary of class-level refactorings with their brief descriptions.}
\label{tab:refactorings}
\end{table*}

To evaluate GPT-4’s ability to perform automated class-level refactoring, we selected a set of techniques from Fowler’s catalog~\cite{fowler2002refactoring} that are directly applicable to object-oriented class design. \tabref{tab:refactorings} lists the refactorings considered in our investigation, each chosen for its relevance to improving internal class structure while preserving external behavior.
These refactorings encompass a broad range of structural improvements, from low-level code cleanups (\eg \textit{Inline Temp}, \textit{Remove Dead Code}) to more architectural transformations (\eg \textit{Replace Method with Method Object}, \textit{Introduce Null Object}). This diversity allows us to assess not only \gpt’s capacity to apply simple edits but also its ability to reason about higher-level design principles when refactoring Python classes.

\subsection{GPT-4o generation process} \label{sub:generation}

Our study focuses on refactoring Python code classes using \gpt. This process is automated through the OpenAI API, which facilitates programmatic interaction with the model. We run the generation process for all tasks in the dataset. The specific model used for the generation is \textit{gpt-4o-mini} (i.e., gpt-4o-mini-2024-07-18) with default parameters (temperature=1.0, top\_p=1.0, max\_tokens=4000). Keeping the default parameters ensures that the experimental scenario closely resembles real-world usage. Since these models do not ensure determinism even with a low temperature, we executed ten iterations for each generation task, and kept, for the analysis, the output of each iteration. 

To ensure that the model consistently performs the generation, we inserted the list of refactorings in the prompt (see \tabref{tab:refactorings}) in order to guide the model during the refactoring. We observed that explicitly specifying the refactorings helps \gpt focus on the target refactorings and avoid applying unrelated refactoring actions. As discussed in previous works~\cite{toufique2023fewshot, suvarna2020review}, detailing the expected output in the prompt helps the model to correctly perform the desired task.

We set the context with the "\textit{system}" prompt in the following textbox. We specify the model’s role and the objective together with an exact output schema and example, hence the response is more deterministic, parseable, and evaluation-ready—consistent with the guidance on structured prompting and format locking~\cite{shinn2023advances}.
We also request the model to display the refactorings performed, as the input method may contain multiple opportunities for refactoring. This list enables us to verify whether the model accurately represents its actions or if it exhibits hallucinations.

The "\textit{user}" prompt is the dataset entry, \ie the code of the class that should be refactored. 

\begin{promptbox}
You are an AI that responds with source code and natural language. You will be given a Python class. Please apply the following refactorings (if applicable) to the given class.\\ Refactorings: \textbf{{', '.join(refactorings)}}

Please return the refactored class and the list of refactorings applied.
The class must be encapsulated in a code block, like this:

\textbf{```python}\\
\textit{print("Hello World!")}\\
\textbf{```}

The list of refactorings must be encapsulated in a text block, like this:

\textbf{```text}\\
\textit{Refactoring 1, Refactoring 2, Refactoring 3}\\
\textbf{```}
\end{promptbox}

\subsection{Analysis Methodology} \label{sub:methodology}

To address RQ1, we leverage the test suite provided with the benchmark to check whether the refactored code maintains the expected behavior. Each task in ClassEval provides a set of tests related to the class, so we run the refactored version generated by \gpt against the available test suite. This analysis highlights \gpt's relative strength or limitations in performing behavior-preserving refactorings.

To address RQ2, we employ four different static analysis tools, \ie Lizard~\cite{lizard}, Pylint~\cite{pylint}, Flake8~\cite{flake8}, and Sonarcloud~\cite{sonarcloud}, to assess the quality of the refactored code compared to the original version. These tools provide extensive reports from different aspects: Lizard measures code complexity (e.g., cyclomatic complexity and function size), Pylint and Flake8 check for style and coding standard violations, while SonarCloud identifies design issues such as code smells, potential bugs, and maintainability problems. This multi-faceted analysis allows us to determine whether \gpt’s refactorings lead to measurable improvements in internal code quality.

To address RQ3, we use the tool  by Scalabrino \etal~\cite{scalabrino2018comprehensive}. Such a tool predicts the readability of source‐code snippets by combining traditional structural features (such as indentation, line length, and spacing) with newly introduced textual features that capture the clarity and consistency of identifiers and comments. Given a piece of code, it extracts these metrics, feeds them into a trained machine-learning model, and outputs a readability score that closely approximates human judgments. The tool outputs a numeric readability score, normalized between zero (hard to read) and one (very easy to read), which represents the predicted probability—based on the combined structural and textual features—that a snippet will be judged readable by humans. We compare the original and refactored versions to assess whether \gpt’s transformations enhance the ease of understanding and maintaining the code.

\section{Results} \label{sec:results}

\begin{table}[ht]
\centering
\caption{Refactoring counts applied by \gpt across all ten iterations}
\label{tab:refactoring_counts}
\begin{tabular}{l r}
\toprule
Refactoring & Count \\
\midrule
Extract Method & 950 \\
Replace Temp with Query & 671 \\
Inline Method & 608 \\
Remove Assignments to Parameters & 273 \\
Replace Magic Number with Symbolic Constant & 137 \\
Self Encapsulate Field & 121 \\
Split Temporary Variable & 101 \\
Consolidate Conditional Expression & 97 \\
Decompose Conditional & 66 \\
Encapsulate Field & 62 \\
Replace Loop with Pipeline & 43 \\
Consolidate Duplicate Conditional Fragments & 36 \\
Remove Dead Code & 22 \\
Replace Method with Method Object & 13 \\
Introduce Explaining Variable & 11 \\
Remove Control Flag & 7 \\
Replace Nested Conditional with Guard Clauses & 7 \\
Rename Variable & 5 \\
Inline Temp & 3 \\
Replace Data Value with Object & 3 \\
Introduce Null Object & 2 \\
Encapsulate Collection & 1 \\
\bottomrule
\end{tabular}

\end{table}

This section presents the outcomes of the study, presenting results of code refactorings generated by \gpt. Table~\ref{tab:refactoring_counts} reports the total counts of refactoring operations applied across all iterations of the experiment. The table shows that the most frequent refactorings are \textit{Extract Method}, \textit{Replace Temp with Query}, and \textit{Inline Method}.

\begin{figure}[t]
  \centering
  \includegraphics[width=\linewidth]{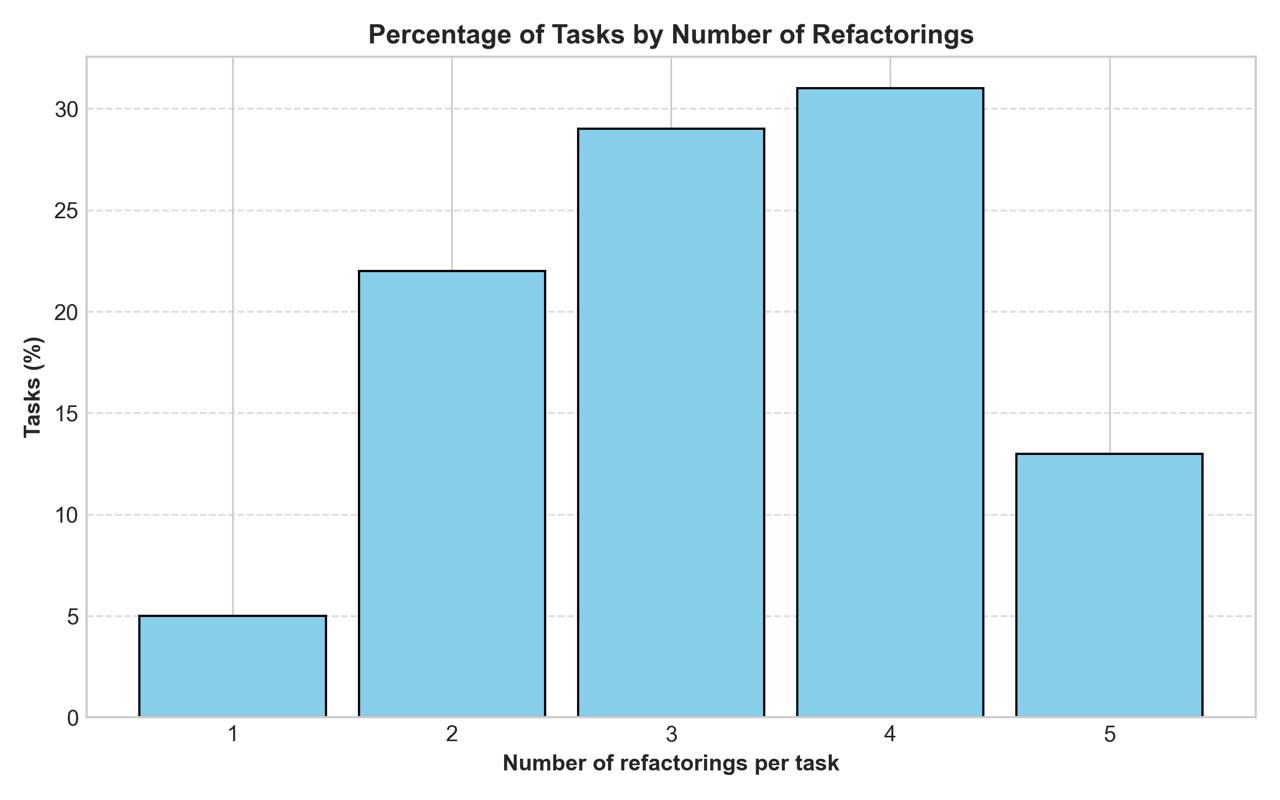}
  \caption{Distribution of the mean values of the number of refactorings per task across 10 runs.}
  \label{fig:ref_means}
\end{figure}

Figure \ref{fig:ref_means} summarizes the distribution of the number of refactorings  applied in the given tasks (a task is a request for the LLM related to an initial version of a code snippet). 
The mean value of the number of refactorings is between one and five per task, with more than 30\% of the tasks with 4 refactorings and between 25-30\% with 3 refactorings. Only a small fraction of tasks sits near the extremes (1 or 5).
Overall, \gpt consistently performs multiple refactorings per task, with most tasks averaging between two and five refactorings. 

Building on this evidence, the following subsections report and discuss results in relation to our three research questions (see Section~\ref{sec:design}): (RQ1) whether refactorings preserve code behavior, (RQ2) the impact of refactorings on code quality assessed with static analysis tools, and (RQ3) the extent to which refactorings improve code readability.

\subsection{RQ1: Preserving Code Behavior}
To address RQ1, we evaluate whether the refactorings generated by the LLM preserve the original program behavior. Since functional correctness is the primary requirement of any refactoring, we rely on the execution of unit tests to assess whether refactorings preserved the original code behavior.

\begin{table}[ht]
\centering
\caption{Evaluation results of LLM-based code refactoring on the \textsc{ClassEval} dataset (10 runs). 
We report average Test Pass Rate (TPR), standard deviation across runs, pass@k, and consistency.}
\label{tab:refactoring-results}
\resizebox{\columnwidth}{!}{%
    \begin{tabular}{cccccc}
    \toprule
    \textbf{Mean TPR} & \textbf{Std TPR} & \textbf{pass@1} & \textbf{pass@5} & \textbf{pass@10} & \textbf{Consistency} \\
    \midrule
    0.844 & 0.01 & 0.86 & 0.88 & 0.89 & 0.8 \\
    \bottomrule
    \end{tabular}
}
\end{table}

Table \ref{tab:refactoring-results} summarizes the correctness of refactorings produced by the LLM on the \textsc{ClassEval} dataset over 10 independent runs. The \textit{Test Pass Rate} (TPR) measures the proportion of tasks whose refactored code passes all unit tests in a single run, while \textit{Std TPR} indicates the variability across runs. The \textit{pass@k} metric reports the percentages of tasks for which at least one out of $k$ runs produced a fully correct refactoring. Finally, Consistency denotes the proportion of tasks that are solved correctly in all runs. 

The results indicate that the model achieves a mean TPR of 0.844 with very low variance (Std TPR = 0.01), suggesting stable performance across runs. Despite the inherent non-determinism of \gpt, the model consistently produces refactorings that preserve program behavior. The \textit{pass@k} values saturate quickly (0.86 at pass@1 and 0.89 at pass@10), implying that additional sampling yields marginal improvements and further supports the model’s stability. According to the consistency metric, 8 out of 10 refactorings pass all tests across all runs, confirming that the majority of generated refactorings maintain the original code behavior. 

Previous studies also employed the pass@k metric to assess the correctness of LLM-generated refactorings. Cordeiro \etal~\cite{cordeiro2024empirical} reported an average of 57.2\% across pass@1, 3, and 5, using StarCoder2 LLM, while Shirafuji \etal~\cite{shirafuji2023refactoring} achieved 91.11\% for pass@1 and 95.68\% for pass@10 using GPT-3.5. Our pass@10 result is in line with the latter, highlighting that \gpt effectively performs behavior-preserving refactorings.

\begin{figure*}[t]
\centering
\noindent
\begin{minipage}[c]{0.48\textwidth} 
\begin{adjustbox}{valign=c}
\begin{lstlisting}[caption={Refactored version}, label={}]
def is_room_booked(self, room_type, name):
    return room_type in self.booked_rooms and name in self.booked_rooms[room_type]

def is_correct_room_number(self, room_type, name, room_number):
    return room_number == self.booked_rooms[room_type][name]

def remove_booking(self, room_type, name):
    self.booked_rooms[room_type].pop(name)

def decrease_booked_room_number(self, room_type, name, room_number):
    self.booked_rooms[room_type][name] -= room_number
        
def check_in(self, room_type, room_number, name):
    if not self.is_room_booked(room_type, name):
        return False
    if self.is_correct_room_number(room_type, name, room_number):
        self.remove_booking(room_type, name)
    else:
        self.decrease_booked_room_number(room_type, name, room_number)
\end{lstlisting}
\end{adjustbox}
\end{minipage}%
\hfill
\begin{minipage}[c]{0.48\textwidth}
\begin{lstlisting}[caption={Canonical version}, label={}]
def check_in(self, room_type, room_number, name):
    # Check if the room of the specified type and number is booked
    if room_type not in self.booked_rooms.keys():
        return False
    if name in self.booked_rooms[room_type]:
        if room_number > self.booked_rooms[room_type][name]:
            return False
        elif room_number == self.booked_rooms[room_type][name]:
            # Check in the room by removing it from the booked_rooms dictionary
            self.booked_rooms[room_type].pop(name)
        else:
            self.booked_rooms[room_type][name] -= room_number
\end{lstlisting}
\end{minipage}
\caption{Comparison of two versions for a snippet in the dataset: on the left, the refactored version by \gpt, while on the right, the canonical solution by ClassEval. Unlike the canonical solution, the refactored version does not pass the tests.}
\label{fig:tests_comparison}
\end{figure*}

Figure~\ref{fig:tests_comparison} shows a comparison between the canonical code and the refactored version by \gpt. On the left, the refactored version of the code, while on the right, the canonical solution proposed by the dataset. The refactored version of the check\_in method replaces the original nested and data-dependent logic with a clearer, modular structure that delegates responsibilities to helper methods. In the canonical code, the function directly accessed and manipulated the \textit{self.booked\_rooms} dictionary, performing multiple conditional checks and updates inline. This made the code harder to read, maintain, and test. In contrast, the refactored version abstracts these operations into dedicated methods, \ie \textit{\url{is\_room\_booked}}, \textit{\url{is\_correct\_room\_number}}, \textit{\url{remove\_booking}}, and \textit{\url{decrease\_booked\_room\_number}}, thereby improving readability and encapsulation.

However, such refactoring has altered the original behavior of the function. The canonical \emph{check\_in} implementation implements a clear three-way guard: if the guest attempts to check in more rooms than booked, it returns False and does not mutate the state. If the number matches, it removes the booking. If the guest checks in fewer rooms than booked, it decrements the booked count. The refactored version streamlines control flow behind helpers, but collapses the three-way logic into a binary check—equal vs not equal—and, in the “not equal” branch, always decrements the booking. Consequently, when \textit{room\_number} > \textit{booked} (\eg guest 1 tries to check in 3 while only 2 are booked, or guest 2 tries 2 while 1 is booked), the refactored code wrongly mutates\textit{ booked\_rooms} (can even go negative) and returns None instead of False. The tests explicitly require that over–check-in attempts return False and leave \textit{booked\_rooms} unchanged, therefore, this behavior causes those assertions to fail. 

The results in Table~\ref{tab:refactoring-results} suggest that the LLM can reliably generate refactorings that preserve the original program behavior in most cases. The rapid saturation of the \textit{pass@k} metric further indicates that the model’s stochasticity has limited impact on correctness, reinforcing the consistency of its outputs. However, the example in Figure~\ref{fig:tests_comparison} reveals that syntactically and structurally sound refactorings may still introduce subtle semantic deviations. In this case, the LLM produced a seemingly cleaner and more modular design, but in doing so altered the control flow and violated a key behavioral constraint. Such errors often stem from the model’s tendency to prioritize readability and abstraction patterns over semantic preservation. This finding highlights an important limitation of LLM-based refactoring: correctness cannot be inferred solely from code quality or structure. Even when the model achieves high aggregate pass rates, individual refactorings may silently compromise behavioral equivalence. Therefore, automated verification through unit tests remains essential to ensure that refactorings maintain functional correctness.

\begin{resultbox}
\textbf{RQ$_1$ summary}: LLM-based refactorings largely preserve program behavior, achieving a mean Test Pass Rate of 0.844 with low variance across runs and high consistency (0.8). Most correct refactorings appear within the first few generations, indicating stable performance. However, occasional semantic deviations show that structural improvements do not guarantee correctness, underscoring the need for systematic test-based verification. 
\end{resultbox}

\subsection{RQ2: Code Quality through Static Analysis Tools}

To address RQ2, we performed a wide analysis using three different static analysis tools, \ie flake8, pylint, and SonarCloud, to assess whether the refactoring proposed by \gpt has improved the overall quality of the code. Firstly, we consider the metric computed by SonarCloud, such as NcLOC (\ie number of non-commenting lines of code), number of statements, cyclomatic complexity, and cognitive complexity. These metrics provide an overview of the changes to the code structure after the refactoring. After that, we examined the smells and warnings detected by all three tools, focusing on the differences in the number of smells and warnings detected after the refactoring.

\begin{figure}
    \centering
    \includegraphics[width=\linewidth]{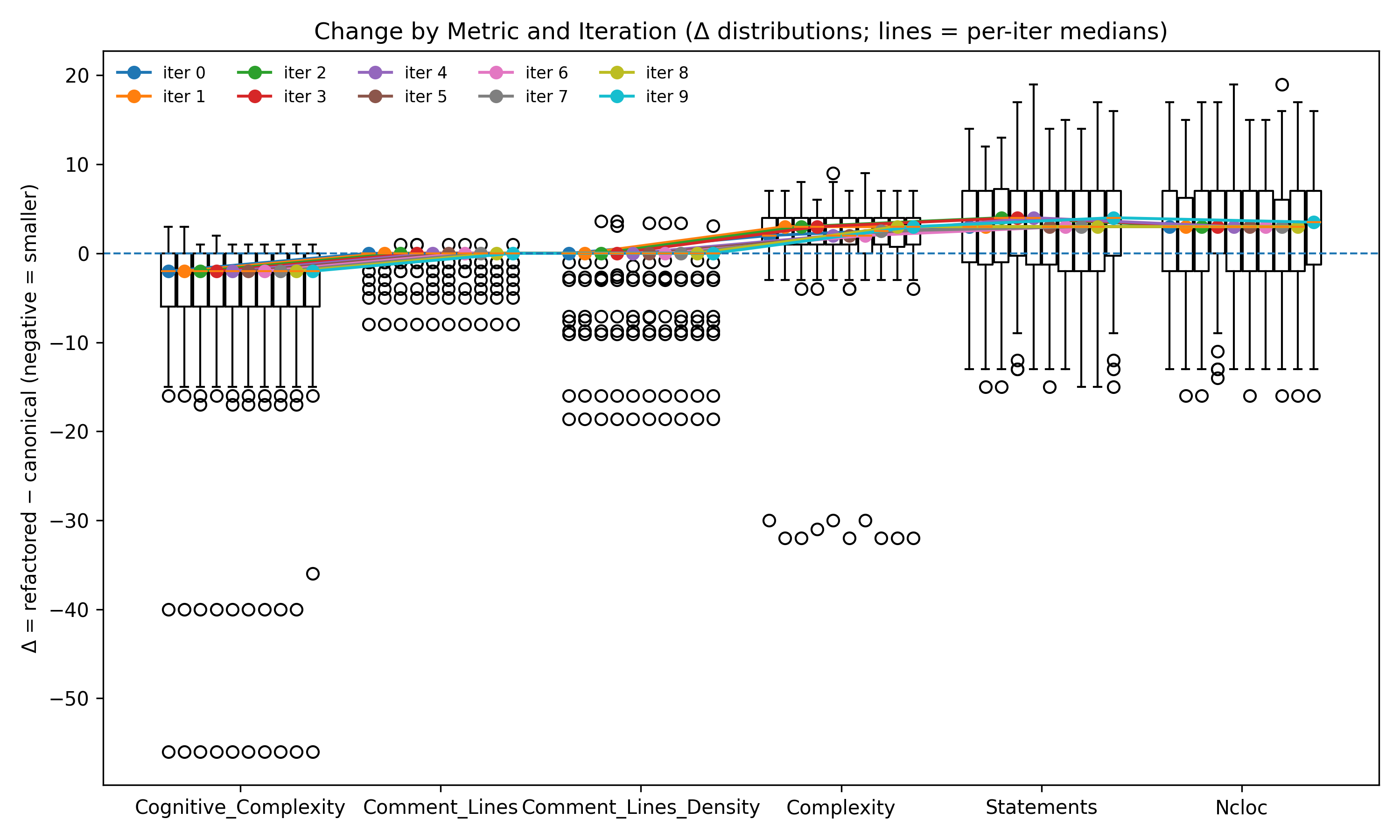}
    \caption{Distribution of metric differences across ten \gpt refactoring runs. Each group of boxes corresponds to a metric, and each color represents a distinct iteration. Negative values indicate smaller or less complex refactored code.}
    \label{fig:metricsboxplot}
\end{figure}

Figure~\ref{fig:metricsboxplot} presents the distribution of absolute changes ($\delta$ = refactored - canonical) for four structural metrics—Cognitive Complexity, Complexity, Statements, and NcLOC (non-commented lines of code), across the ten \gpt refactoring runs. Each box summarizes how the metric values in the refactored code differ from their corresponding canonical implementations, while the colored lines indicate the median values per iteration. Negative values represent improvements, as they correspond to more compact or less complex refactorings. The overall pattern reveals that \gpt tends to generate code of comparable or slightly lower complexity than the canonical version, particularly for Cognitive Complexity, where several iterations show consistent negative median shifts. The other metrics (Complexity, Statements, and NcLOC) exhibit smaller variations around zero, indicating that \gpt generally preserves structural size and statement count rather than performing aggressive simplifications. The narrow spread of medians across iterations suggests stable behavior of the model in this respect, with outliers corresponding to isolated cases of substantially shorter or more complex refactorings.

\begin{figure}
    \centering
    \includegraphics[width=\linewidth]{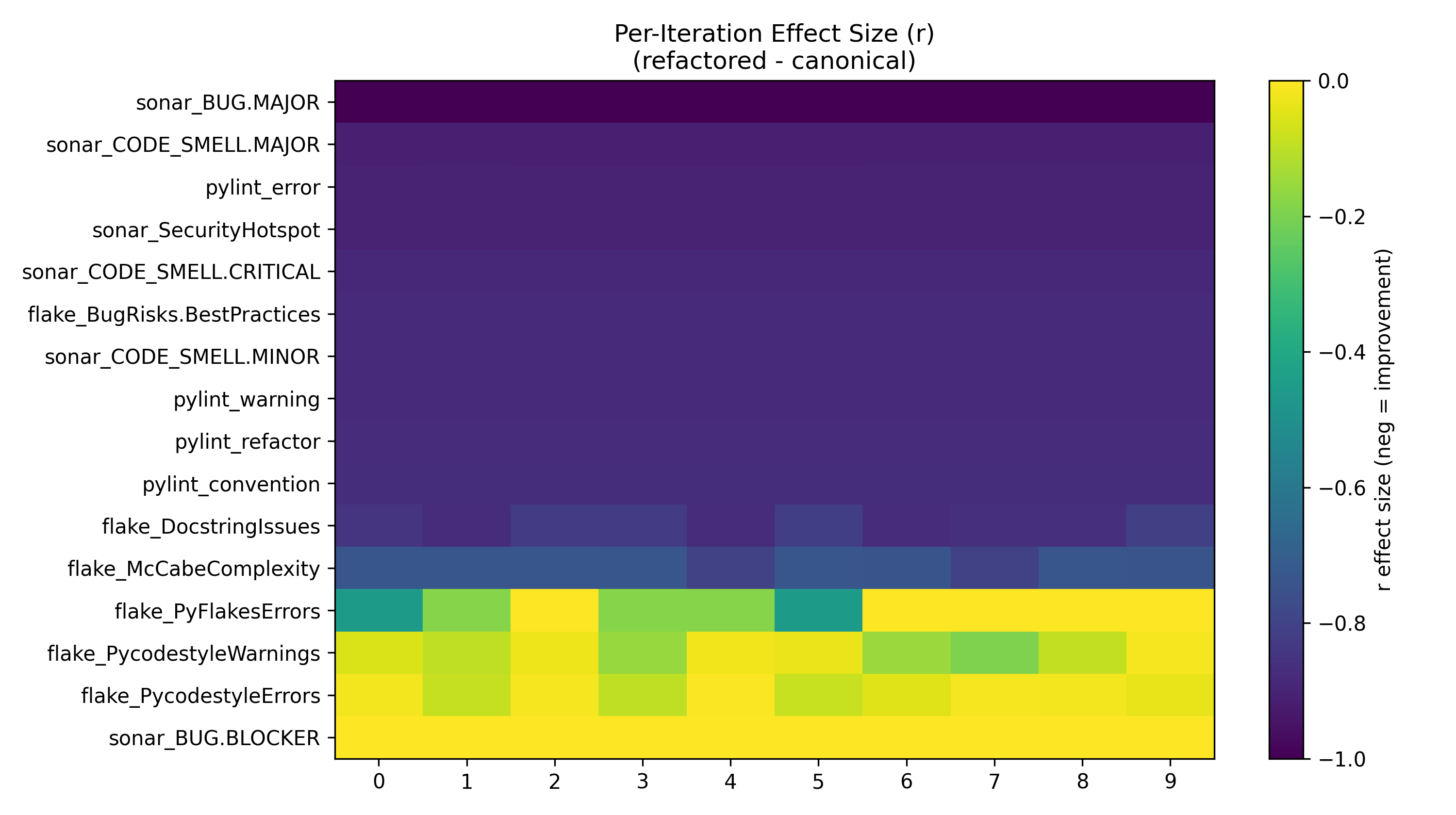}
    \caption{Per-iteration Wilcoxon effect sizes (r) comparing refactored and canonical code across all static analysis metrics. Each cell represents the effect size for a specific metric (row) and \gpt refactoring run (column). Negative values (darker colors) indicate improvements, meaning the refactored code exhibits fewer detected issues.}
    \label{fig:warningsheatmap}
\end{figure}

Figure~\ref{fig:warningsheatmap} illustrates the per-iteration distribution of the Wilcoxon effect sizes (r) for each static analysis metric, comparing the refactored code generated by \gpt against the canonical implementations in the ClassEval dataset. Each row corresponds to a specific metric from Flake8, Pylint, or SonarCloud, while each column represents one of the ten refactoring runs. The color intensity encodes the magnitude and direction of the effect size, where negative values (darker shades) indicate an improvement, \ie the refactored version exhibiting fewer reported issues than the canonical one. Figure~\ref{fig:warningsheatmap} shows that improvements are mainly in stylistic metrics such as flake\_PycodestyleErrors, flake\_PycodestyleWarnings, and flake\_PyFlakesErrors, with several iterations yielding moderate to strong negative effect sizes. In contrast, metrics related to higher-level quality dimensions—such as code smells and bug risks from SonarCloud, or Pylint warnings and refactors—exhibit small or negligible effects, suggesting limited impact of \gpt refactoring on deeper structural or semantic aspects of code quality. Overall, the heatmap highlights a consistent trend: \gpt tends to improve surface-level code cleanliness while leaving more complex quality indicators largely unaffected across iterations. This outcome aligns with prior evidence suggesting that refactoring actions are not necessarily triggered by problematic metric profiles or the presence of code smells, but are instead guided by a broader intent to enhance maintainability~\cite{bavota2015experimental}. The table with the metrics and clear data can be found in the replication package~\cite{replication}.

\begin{figure*}[t]
\centering
\noindent
\begin{minipage}[c]{0.48\textwidth} 
\begin{adjustbox}{valign=c}
\begin{lstlisting}[caption={Refactored version}, label={}]
@staticmethod
def _first_positive(array):
    for value in array:
        if value > 0:
            return value
    return 0.0
    
@staticmethod
def _calculate_mrr_for_tuple(data):
    (sub_list, total_num) = data
    sub_list = np.array(sub_list)
    if total_num == 0:
        return 0.0, [0.0]

    ranking_array = 1.0 / (np.arange(len(sub_list)) + 1)
    mr_np = sub_list * ranking_array

    mr = MetricsCalculator2._first_positive(mr_np)
    return mr, [mr]
\end{lstlisting}
\end{adjustbox}
\end{minipage}%
\hfill
\begin{minipage}[c]{0.48\textwidth}
\begin{lstlisting}[caption={Canonical version}, label={}]
if total_num == 0:
    return 0.0, [0.0]
else:
    ranking_array = 1.0 / (np.array(list(range(len(sub_list)))) + 1)
    mr_np = sub_list * ranking_array

    mr = 0.0
    for team in mr_np:
        if team > 0:
            mr = team
            break
    return mr, [mr]
\end{lstlisting}
\end{minipage}
\caption{Comparison between the canonical (right) and refactored (left) implementations of a fragment from ClassEval\_57 task. This refactoring eliminates redundant logic, simplifies control flow, and contributes to the reduction of Flake8 warnings observed in the static analysis.}
\label{fig:warningsComparisons}
\end{figure*}

Figure~\ref{fig:warningsComparisons} shows two code fragments extracted from ClassEval\_57 task. On the left, the figure shows the refactored version by \gpt, while on the right, the canonical solution from the dataset. The refactored version exhibits a clear improvement in static analysis outcomes, showing fewer Flake8 warnings compared to the canonical implementation (10 vs. 22 issues). Most notably, all E721 (\ie Do not compare types, use isinstance()) and CCR001 (\ie Cognitive complexity is too high) violations were removed, and the number of E501 (\ie Line too long) line-length errors was halved, confirming that the refactoring improved both stylistic and structural quality without introducing new problems. Here, the refactoring keeps the external API and return formats intact but streamlines the internals: (i) \textit{a single\_validate\_input} gate replaces repeated type checks; (ii) tuple/list cases for both \textit{mrr} and \textit{map} are funneled into small helpers (\textit{\_calculate\_*\_for\_tuple} / \textit{\_calculate\_*\_for\_list}) that remove duplicated logic; (iii) imperative loops are replaced by compact utilities, \eg\textit{\_first\_positive}, and NumPy idioms (\textit{np.arange} instead of \textit{np.array(list(range(...))))}, reducing branching; and (iv) operations like building the right‐ranking sequence are encapsulated in \textit{\_calculate\_right\_ranking}. Net effect: less code duplication, lower cognitive/branch complexity, and clearer control flow, while preserving behavior on empty inputs and zero totals.

\begin{resultbox}
\textbf{RQ$_2$ summary:}
\gpt-based refactorings yield moderate quality improvements, mainly reducing stylistic and formatting issues detected by Flake8, while leaving structural and semantic properties largely unchanged. Minor reductions in cognitive complexity suggest slightly clearer control flow, yet design metrics from SonarCloud remain stable. Overall, \gpt enhances surface-level code cleanliness without significantly affecting underlying quality attributes.
\end{resultbox}

\subsection{RQ3: Code Readabilty Assessment}

To address RQ3, we employed the tool of Scalabrino \etal~\cite{scalabrino2018comprehensive} to compute a readability score (in [0, 1]) for both the canonical and the refactored versions of each task. The objective was to verify whether the refactorings generated by \gpt improved code readability.

\begin{figure}
    \centering
    \includegraphics[width=0.9\linewidth]{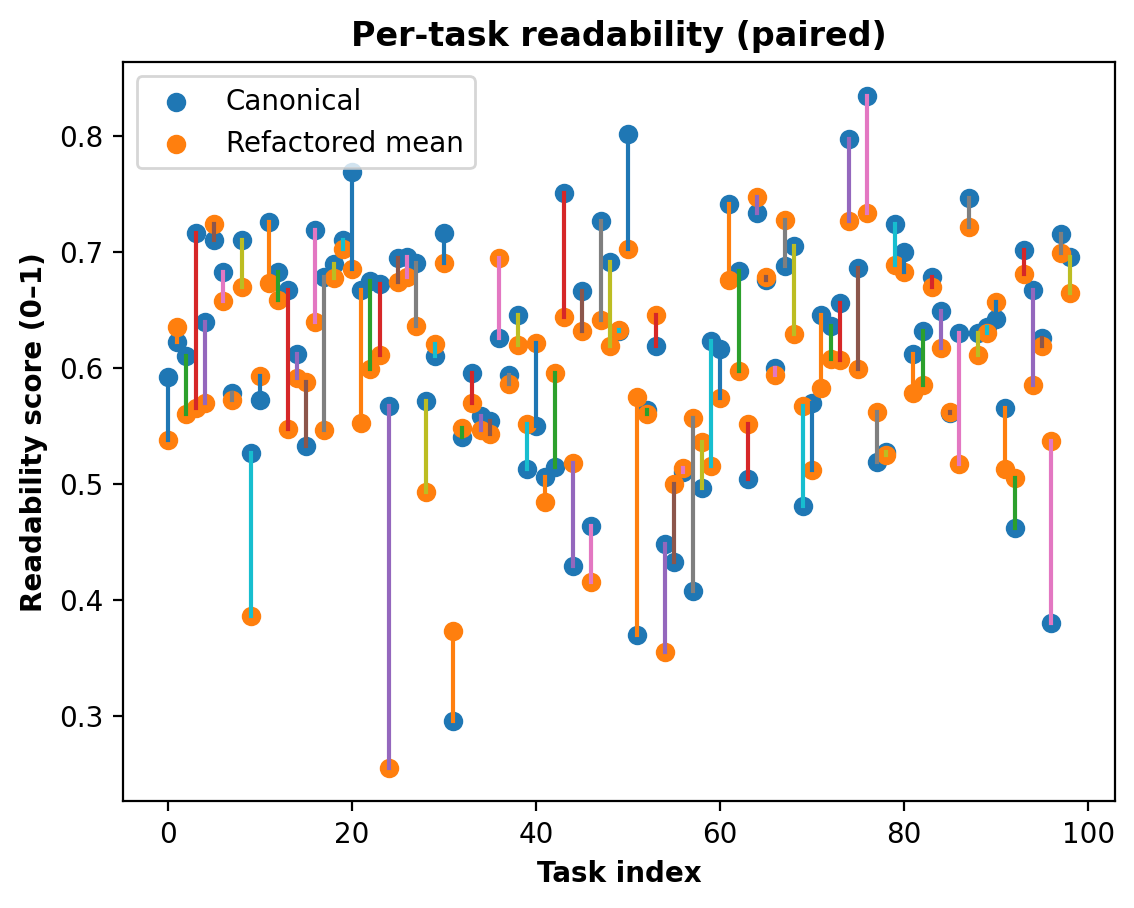}
    \caption{The paired scatter plot of readability scores between the refactored and canonical versions of the same task. Blue canonical solution scores, while orange for refactoring scores.}
    \label{fig:pairedread}
\end{figure}

\begin{table}[ht]
\centering
\caption{Statistical analysis of readability before and after refactoring.}
\label{tab:readability_stats}
\begin{tabular}{lcc}
\toprule
\textbf{Metric} & \textbf{Value} \\
\midrule
Before mean (SD) & 0.6199 (0.1002) \\
After mean (SD) & 0.5959 (0.0847) \\
$\Delta$ mean (SD) & $-0.0240$ (0.0690) \\
$\Delta$ 95\% CI & $[-0.0378, -0.0102]$ \\
\% improved & 29.3\% \\
\% worsened & 70.7\% \\
Wilcoxon $W$ & 1350.000 \\
Wilcoxon $p$-value & 1.000 \\
Cliff's $\delta$ & $-0.183$ \\
\bottomrule
\end{tabular}
\end{table}

Table \ref{tab:readability_stats} reports the descriptive and inferential statistics for the comparison between the canonical and refactored code versions. On average, the readability score decreased from 0.6199 (SD = 0.1002) to 0.5959 (SD = 0.0847), corresponding to a mean difference of $-0.024$ (95\% CI [$-0.0378$, $-0.0102$]). Readability improved in 29.3\% of the tasks and worsened in 70.7\%, indicating that refactoring led to a deterioration of perceived readability in the majority of cases. The Wilcoxon signed-rank test yielded $W = 1350$ and $p = 1.000$ (one-sided, ref $>$ canon), suggesting no statistical evidence that the refactored versions were more readable; on the contrary, the direction of the differences consistently favored the canonical implementations. Consistently, Cliff’s $\delta = -0.183$ indicates a small but systematic negative effect, with the negative sign denoting that the canonical code tends to achieve higher readability scores than the refactored one.

Figure \ref{fig:pairedread} visualizes the readability scores of the canonical and refactored versions for each of the 99 tasks. Each vertical segment connects the two scores for the same task, highlighting whether the LLM-generated refactoring increased or decreased readability. In most cases, the orange marker (refactored mean) lies below the blue marker (canonical), indicating a decrease in readability. Only a minority of tasks show upward segments, corresponding to improvements. The spread of values suggests that readability varies substantially across tasks, but the overall pattern is consistent with the statistical results in Table \ref{tab:readability_stats}: the majority of refactorings reduce readability, and the mean shift is negative. The fact that the refactored means cluster more tightly than the canonical scores also reflects the aggregation of ten LLM runs per task, which smooths random fluctuations but does not reverse the overall downward trend. In summary, the paired plot visually reinforces that readability degradation after LLM-based refactoring is systematic rather than incidental.

\begin{figure*}[t]
\centering
\noindent
\begin{minipage}[c]{0.48\textwidth} 
\begin{adjustbox}{valign=c}
\begin{lstlisting}[caption={Refactored version}, label={}]
class ComplexCalculator:
    def __init__(self):
        pass

    @staticmethod
    def add(c1, c2):
        return ComplexCalculator._calculate(c1, c2, lambda r1, r2, i1, i2: (r1 + r2, i1 + i2))

    @staticmethod
    def subtract(c1, c2):
        return ComplexCalculator._calculate(c1, c2, lambda r1, r2, i1, i2: (r1 - r2, i1 - i2))

    @staticmethod
    def multiply(c1, c2):
        return ComplexCalculator._calculate(c1, c2, lambda r1, r2, i1, i2: (r1 * r2 - i1 * i2, r1 * i2 + i1 * r2))

    @staticmethod
    def divide(c1, c2):
        denominator = c2.real**2 + c2.imag**2
        return ComplexCalculator._calculate(c1, c2, lambda r1, r2, i1, i2: (
            (r1 * r2 + i1 * i2) / denominator,
            (i1 * r2 - r1 * i2) / denominator
        ))

    @staticmethod
    def _calculate(c1, c2, operation):
        real = operation(c1.real, c2.real, c1.imag, c2.imag)
        return complex(real[0], real[1])
\end{lstlisting}
\end{adjustbox}
\end{minipage}%
\hfill
\begin{minipage}[c]{0.48\textwidth}
\begin{lstlisting}[caption={Canonical version}, label={}]
class ComplexCalculator:
    def __init__(self):
        pass

    @staticmethod
    def add(c1, c2):
        real = c1.real + c2.real
        imaginary = c1.imag + c2.imag
        answer = complex(real, imaginary)
        return answer
    
    @staticmethod
    def subtract(c1, c2):
        real = c1.real - c2.real
        imaginary = c1.imag - c2.imag
        return complex(real, imaginary)
    
    @staticmethod
    def multiply(c1, c2):
        real = c1.real * c2.real - c1.imag * c2.imag
        imaginary = c1.real * c2.imag + c1.imag * c2.real
        return complex(real, imaginary)
    
    @staticmethod
    def divide(c1, c2):
        denominator = c2.real**2 + c2.imag**2
        real = (c1.real * c2.real + c1.imag * c2.imag) / denominator
        imaginary = (c1.imag * c2.real - c1.real * c2.imag) / denominator
        return complex(real, imaginary)
\end{lstlisting}
\end{minipage}
\caption{Comparison of two implementations of a task in the dataset: on the left, the refactored version by \gpt, while on the right, the canonical solution proposed by ClassEval. The refactored version scored a readability of 0.24, while the canonical solution scored 0.56.}
\label{fig:readability_comparison}
\end{figure*}

Figure~\ref{fig:readability_comparison} compares the canonical and refactored implementations of the \texttt{ComplexCalculator} class. Although both versions are functionally equivalent, the refactored code exhibits lower readability according to the Scalabrino \textit{\etal}~\cite{scalabrino2018comprehensive} model. In the refactored version, the LLM introduces a generic private method, \texttt{\_calculate}, which abstracts the arithmetic logic through higher-order lambda expressions passed to the operation methods (\texttt{add}, \texttt{subtract}, etc.). This design reduces code duplication but substantially increases the cognitive load: understanding each arithmetic operation now requires mentally unpacking a lambda expression and tracing its parameters within a shared helper function. Moreover, the use of inline anonymous functions and tuple unpacking reduces the explicitness of intermediate computations, which is clearly visible in the canonical version through the use of named variables (\texttt{real}, \texttt{imaginary}, \texttt{denominator}). The canonical implementation, by contrast, follows a straightforward, step-by-step structure that mirrors the mathematical formulation of complex number operations. Each step is explicit and self-contained, enhancing readability and maintainability, even when it repeats similar patterns across methods. Consequently, while the refactoring improves structural conciseness, it sacrifices clarity and local transparency—two aspects heavily weighted in readability metrics and human comprehension. This example illustrates a recurring behavior observed in our dataset: the LLM tends to over-abstract simple code fragments, prioritizing syntactic compactness over expressive clarity, leading to a measurable reduction in readability. However, readability metrics can sometimes be misleading, as they do not necessarily correlate with actual code understandability~\cite{scalabrino2017automatically}.

\begin{resultbox}
\textbf{RQ$_3$ summary:}
Readability analysis reveals that \gpt-based refactorings generally reduce code readability. The average readability score decreased from 0.62 to 0.60, with only about 30\% of tasks showing improvement. However, such an observed difference is not statistically significant. Although the refactorings often remove redundancy and increase structural conciseness, they tend to introduce abstractions (\eg helper methods or lambdas) that obscure control flow and variable intent. This behavior results in code that is more compact but less transparent, confirming that \gpt prioritizes structural optimization over human readability.
\end{resultbox}

\section{Threats to Validity} \label{sec:threats}

Threats to \emph{construct validity} concern the relationship between theory and observation. 
For RQ$_1$, we relied on the test suites provided in the \textsc{ClassEval} benchmark to evaluate behavioral preservation. Although this dataset is widely used and reports a branch-level coverage of 98.2\%~\cite{xueying2023classeval}, the available test cases may not fully cover all possible program behaviors, limiting the completeness of our assessment.
For RQ$_2$, we employed standard structural metrics (e.g., LOC, cyclomatic complexity) and warnings from widely adopted static analysis tools (Flake8, Pylint, SonarCloud) to assess code quality. Different tools or configurations might yield slightly different results, but our selection reflects common industry practice.
For RQ$_3$, we used the readability model proposed by Scalabrino \textit{\etal}~\cite{scalabrino2018comprehensive}, which, despite its limitations, is the most widely used and validated approach for automated readability estimation in source code. Moreover, readability should not be conflated with understandability: as shown by Scalabrino \textit{\etal}~\cite{scalabrino2017automatically}, none of the existing readability or complexity metrics exhibited a strong correlation with developers’ actual or perceived code understanding, indicating that readability captures only one facet of understandability.

Threats to \emph{internal validity} relate to factors internal to our study that could have influenced the observed results.
The main threat concerns prompt design, as variations in prompt formulation may affect LLM behavior. To mitigate this, we followed prompt design guidelines from prior work on code generation~\cite{shinn2023advances}.
Another factor is the inherent non-determinism of LLMs. We mitigated this by executing ten independent runs per task and aggregating results to reduce stochastic variability.

Threats to \emph{conclusion validity} concern the soundness of the inferences drawn from our data. We applied appropriate statistical analyses, including non-parametric tests, $p$-value adjustments, and effect size computations, to ensure that our conclusions are statistically robust and not due to random chance.

Threats to \emph{external validity} concern the generalizability of our findings.
Our experiments were conducted on the \textsc{ClassEval} dataset, which, although state-of-the-art for code generation, was not originally designed for refactoring. Consequently, results may not generalize to other software systems or larger, real-world projects. Nonetheless, our setup reflects a realistic scenario in which a developer requests an LLM to refactor existing code that may or may not require modification.
The choice of language (Python) may also limit generalization, as other languages differ in syntax, idioms, and tooling support, which could influence the effectiveness of LLM-based refactoring.
Finally, the proprietary nature of the LLM introduces a potential threat to replicability: future model updates or API changes could result in slightly different outputs, even under the same experimental setup. Future work could compare results across programming languages, open-source models, and multiple API versions to better assess the reproducibility and generalizability of the findings.

\section{Conclusion} \label{sec:conclusion}

This study presented a comprehensive empirical investigation into LLM-driven code refactoring using \gpt. By leveraging the \classeval benchmark and a wide spectrum of class-level refactorings from Fowler’s catalog, we systematically examined the ability of LLMs to perform meaningful, behavior-preserving refactorings. Through a multi-perspective evaluation, encompassing correctness, code quality, and readability, we provided a holistic view of the current strengths and limitations of LLM-assisted refactoring.

Our findings indicate that \gpt can effectively perform a variety of refactorings while maintaining behavioral correctness in many cases. However, improvements in static analysis metrics and readability are not always consistent, revealing that LLMs may optimize structural aspects at the expense of code clarity or introduce stylistic inconsistencies. 

These results underscore both the promise and the current immaturity of LLM-based refactoring. While LLMs demonstrate clear potential to support developers in automating complex transformations, ensuring reliability and systematic quality improvement remains an open challenge. Future research should focus on hybrid approaches that integrate LLM reasoning with program analysis and testing, as well as the definition of standardized benchmarks for multifaceted quality evaluation.

\balance
\bibliographystyle{ACM-Reference-Format}
\bibliography{main}

\end{document}